\documentclass[a4paper,12pt]{article}
\usepackage{amssymb,amsmath}

\newcommand{\e}{{\rm e}}

\newcommand{\gap}{\vspace{.2in}}

\title{ On the Spectrum of a Noncommutative
Formulation of the D=11 Supermembrane with Winding.}
\author{\text{ M.P. Garc\'{\i}a del Moral$^a$\dag\quad and A. Restuccia \dag \ddag}\\
\setlength{\baselineskip}{6pt}
\ {\small \dag Phys. Dept. Universidad S. Bol\'{\i}var, Caracas.}\\
{\small \ddag Dept. of Mathematics King's College, London.}\\
\footnote{\em e-mail: mgarcia@fis.usb.ve, arestu@usb.ve}
}
\date{30 March 2001}

\begin{document}

\maketitle
\ \vskip -4in {\em \
\parbox[t]{1.1\linewidth}{\small
\rightline{KCL-MTH-01-06} } }

\vskip 3.5in

\begin{abstract}
A regularized model of the double compactified D=11 supermembrane with nontrivial winding
 in terms of SU(N) valued maps is obtained. The condition of nontrivial winding
 is described in terms of a nontrivial line bundle introduced in the formulation of
 the compactified supermembrane. The
multivalued geometrical objects of the model related to the nontrivial wrapping
are described in terms of a SU(N) geometrical object  which in the $ N\to \infty$
limit, converges to the symplectic connection related to the area preserving
diffeomorphisms of the recently obtained non-commutative description of the compactified D=11
supermembrane  \cite{Martin},\cite{Ovalle1}.

The SU(N) regularized canonical lagrangian is explicitly obtained. In the $ N\to \infty$
limit it converges to the lagrangian in \cite{Martin}\cite{Ovalle1} subject
 to the nontrivial winding condition. The spectrum
of the hamiltonian of the double compactified D=11 supermembrane is discussed.
 Generically, it contains local string like spikes with zero energy.
 However the sector of the theory corresponding to a
principle bundle characterized by the winding number $n \neq 0$,
described by the SU(N) model we propose, is shown to
have no local string-like spikes and hence the spectrum of this sector
should be discrete.
 \end{abstract}

\setlength{\parindent}{0cm}

\setlength{\baselineskip}{15pt}
 {\em Keywords: supermembrane, matrix models, SU(N) regularization,\\
                noncommutative geometry}

\setlength{\parindent}{.6cm}

\setlength{\baselineskip}{15pt}

\pagebreak

\section{Introduction}

The matrix model for bosonic membranes
was first introduced in \cite{Goldstone}, and its study was extended to the
supersymmetric case in \cite{Hoppe}. In that work, it was shown that the
supermembrane theory could also be understood as a supersymmetric  gauge theory
of the infinite group of area preserving diffeomorphisms which appeared as a residual symmetry of
the supermembrane in the light cone gauge. They also found
a satisfactory SU(N) regularization
of the model.  The spectrum of the
quantized model was found to be continuous \cite{Luscher & Nicolai}, and  was afterwards interpreted in
terms of a multiparticle theory  \cite{Russo}.

The D=11 supermembrane with winding was first analyzed in  \cite{Peeters & Plefka} in
terms of multivalued maps from the world volume to the target space.
Part of their study
 was based on a previous work on the area preserving diffeomorphisms in
 \cite{Marquard}.
In \cite{Peeters & Plefka} the
hamiltonian of the theory was explicitly obtained. Its analysis in terms of a
finite $N$ regularization was performed, with the
conclusion that the SU(N) regularization of the model, which was essential in
the analysis of the spectrum of the non-compactified supermembrane in a D=11
Minkowski target space  \cite{Luscher & Nicolai}, was not possible because the
structure constants associated to the presence of the non-exact modes did
not fit in an SU(N) description of the model. It was also argued in
 \cite{Peeters & Plefka}
that the spectrum of the compactified D=11 supermembrane should
also be continuous since the instability, caused by the
string-like spikes, is also present in the compactified case. The
string-like spikes are singular physical configurations (the
determinant of the induced metric is zero at some points or open sets
of the worldvolume) which may even change
the topology of the world volume without changing the energy of
the system. Together with the supersymmetry they render the
spectrum of the D=11 supermembrane continuous from zero to
infinity. A complete analysis of the spectrum for the compactified
case similar to the one in \cite{Luscher & Nicolai}, for the non
compactified case, has not been yet presented.

 The analysis of the compactified D-brane was first approached from the matrix
model point of view in \cite{Taylor}. The matrix models  \cite{Banks},
 \cite{Ishibashi} describe the dynamics of the membranes in the Light
 Cone Gauge in
the approximation of finite number of oscillations modes. They
provide an equivalent description, to the one in \cite{Peeters & Plefka} of the
supermembrane in terms of D0-branes. The formulation
of compactified D-branes in \cite{Taylor} was done by considering
the universal covering of the compactified target space. In that
simply connected space the matrix model may be directly formulated
in terms of the infinite set of the copies of the D-brane system
restricted by the symmetry generated by the covering group. An
interesting result was obtained in \cite{ Connes}. It was shown
that the matrix model on a noncommutative torus is equivalent to
M-theory compactified in a constant antisymmetric background field.
The non commutative geometry of the supermembrane in terms of
matrix models was also described in several papers, see \cite{W. Taylor}
 and references therein, in particular
in \cite{Floratos} and \cite{Takata}. In \cite{Torrealba},
\cite{Ovalle} the analysis of the compactified D=11 supermembrane
was performed following the original description \cite{Hoppe} but
the analysis emphasizes the global structure associated to the non
trivial wrapping of the supermembrane in terms of an associated
principal bundle which is naturally constructed from the
non-trivial central charge of the supersymmetric algebra. This
analysis is best performed in the dual formulation of the theory.
The double compactified D=11 supermembrane dual directly
introduces the connection 1-form associated to the nontrivial
principal bundle. In the formulation in
 \cite{Martin},\cite{Ovalle1} the canonical lagrangian is expressed as a noncommutative gauge theory. The geometrical meaning of the non commutativity was
 explained in that work in terms of symplectic fibrations over the world
 volume. The symplectomorphisms on the fibers are generated by the area
 preserving diffeomorphisms on the world volume.

 In this paper we present a SU(N) regularization of that formulation. All the
 multivalued objects related to the non trivial wrapping are handled by the
 connections 1-form. If the theory is restricted to a principle bundle
 characterized by the winding number $n$ , any connection on that bundle may
 be expressed in terms of a fix one $\widehat{\Pi}$ plus a uniform
 1-form  $\mathcal{A}$:
 \begin{equation}\label{eqn -1}
 \begin{aligned}
\widehat{\mathcal{A}}= \widehat{\Pi}+\mathcal{A}
\end{aligned}
\end{equation}
We consider $\widehat{\Pi}$ to be the connection 1-form which minimizes
the Hamiltonian of the double compactified Supermembrane. Although  $\mathcal{A}$
is a uniform 1-form it has a transformation law, under the gauge symmetry of
the theory,
\begin{equation}\label{eqn0}
\begin{aligned}
\mathcal{A}\to \mathcal{A} + d\epsilon +
  \{\mathcal{A},\epsilon\}= \mathcal{A}
  +\mathcal{D}\mathcal{\epsilon}
\end{aligned}
\end{equation}
corresponding to a symplectic connection preserving the symplectic
structure of the fibers under holonomies. Its regularization in
terms of SU(N) valued objects has consequently a very different
behaviour compared to the geometrical objects in the other
approaches.

\section{The Hamiltonian of the compactified D=11 Supermembrane.}

In this section we describe the Hamiltonian of the compactified D=11
supermembrane on $(R^{9}\times S^{1} \times S^{1} ) $. It seems that the best
approach, from a global point of view, is to consider its dual
formulation since as discussed previously the global features are geometrically
handled in terms of a  connection 1-form over a non-trivial principle bundle
on the world volume which is intrinsically introduced in the formulation.

The Hamiltonian for the double compactified D=11
supermembrane was obtained in \cite{Martin},\cite{Ovalle1} starting from the lagrangian
formulation of the D=11 Supermembrane. It was important to follow step by step
the dualization procedure in order to show that the non-trivial winding of the
supermembrane was indeed described by the nontrivial bundle over which the
gauge field, dual to the compactified coordinates, is defined. Having that
geometrical structure one may introduce in an intrinsic
way a symplectic structure on the world volume. One finally may
formulate the double compactified D=11 supermembrane as a symplectic
noncommutative gauge theory \cite{Martin}, \cite{Ovalle1}. The final form of the Hamiltonian
is
\begin{equation}\label{eq1}
\begin{aligned}
H= &\int_{\Sigma} \frac{1}{2\sqrt{W}}[ (P^{m})^{2}+(\Pi_{r})^{2}+
 1/2W\{X^{m},X^{n}\}^{2}
+W(\mathcal{D}_r X^{m})^{2}+1/2W(\mathcal{F}_{rs})^{2}]\\&+
\int_{\Sigma}[1/8\sqrt{W}n^{2}
-\Lambda(\mathcal{D}_{r}\Pi^{r}+\{X^{m},P_{m}\})]-\frac{1}{4}\int_{\Sigma}\sqrt{W}n ^{*}\mathcal{F}
\end{aligned}
\end{equation}

together with its supersymmetric extension
\begin{equation}\label{a}
\begin{aligned}
\int_{\Sigma} \sqrt{W} [- \overline{\theta}\Gamma_{-} \Gamma_{r} \mathcal{D}_{r}\theta +
 \overline{\theta}\Gamma_{-} \Gamma_{m}\{X^{m},\theta\} +
 \Lambda \{ \overline{\theta}\Gamma_{-},\theta\}]
\end{aligned}
\end{equation}

in terms of the original Majorana spinors of the D=11 formulation, which may be
decomposed in terms of a complex 8-component spinor of $ SO(7)\times U(1)$.

$m=1,...7$ are the indexes denoting the scalar fields once
 the supermembrane is formulated in the light cone gauge.

 $r,s=1,2$ are the indexes related to the two compactified directions of the
 target space.
Where $\Sigma$ is the spatial part of the world volume which is assumed to be
closed Riemann surface of topology $g$.

$P_M$ and $\Pi_{r}$ are the conjugate momenta to $X^M$ and the
connection 1-form $\mathcal{A}_{r}$ respectively. The covariant derivative is
\begin{equation}\label{b}
\begin{aligned}
\mathcal{D}_{r}= D_{r} + \{\mathcal{A}_{r},\}
\end{aligned}
\end{equation}
and the field strength
\begin{equation}\label{c}
\begin{aligned}
\mathcal{F}_{rs} =D_{r}\mathcal{A}_{s}-
D_{s}\mathcal{A}_{r}+ \{\mathcal{A}_{r},\mathcal{A}_{s}\}
\end{aligned}
\end{equation}

The bracket $\{ , \}$ is defined as
\begin{equation} \label{eq3}
\begin{aligned}
\{\ast,\diamond\}= \frac{2\epsilon^{sr}}{n}(D_{r}\ast)(D_{s}\diamond)
\end{aligned}
\end{equation}

where $n$ denotes the integer which characterizes the non trivial principle
bundle under consideration. $D_{r}$ is a tangent space derivative

\begin{equation} \label{eq3'}
\begin{aligned}
D_{r}\diamond=\frac{\widehat{\Pi}^{a}_{r}\partial_{a}\diamond}{\sqrt{W}}=
\{\widehat{\Pi}_{r},\diamond\}
\qquad \qquad r,s=1,2:a=1,2.
\end{aligned}
\end{equation}

where $\partial_{a}$ denotes derivatives with respect to the local
coordinates of the world volume while
$\widehat{\Pi}^a_{r}=\epsilon^{au}\partial_{u}\Pi_{r}$ is a
zwei-vein defined from the minimal solution of the Hamiltonian of
the theory. It satisfies
\begin{equation}\label{eq3'''}
\begin{aligned}
 \epsilon^{rs}\widehat{\Pi}^a_{r}
 \widehat{\Pi}^b_{s}\epsilon_{ab}=n\sqrt{W}
\end{aligned}
\end{equation}
equivalently

\begin{equation}\label{eq3''''}
\begin{aligned}
 \{\widehat{\Pi}_{r},
 \widehat{\Pi}_{s}\}=1/2n\epsilon_{sr}
\end{aligned}
\end{equation}

 We consider now an expansion of the geometrical objects in the formulation in
 terms of an orthonormal basis in the space $L^{2}$ or functions over the world
 volume. They are uniform functions over the manifold. The $X^{M}, P_{M}$ may
 be expressed in the standard way since they are uniform maps from the world
 volume $\Sigma$ to the target space,

\begin{equation}\label{eq3''}
\begin{aligned}
X^m(\sigma^{1},\sigma^{2},\tau)= & \sum
X^{mA}(\tau)Y_{A}(\sigma^{1},\sigma^{2})\\
P_m(\sigma^{1},\sigma^{2},\tau)=& \sum \sqrt{W}
P^{A}_{m}(\tau)Y_{A}(\sigma^{1},\sigma^{2})
\end{aligned}
\end{equation}
The multivalued maps defining the non-trivial winding of the membrane are now
expressed in terms of the connection and its conjugate momenta. In this sector
one has performed the following decomposition
\begin{equation}\label{eq4}
\begin{aligned}
\widehat{\mathcal{A}}_{s} \to \widehat{\Pi}_{s} + \mathcal{A}_{s}
\end{aligned}
\end{equation}

where $\widehat{\Pi}_{s}$ is the connection 1-form which
minimizes the Hamiltonian of the theory. It is a connection on a
nontrivial bundle characterized by the integer $n\neq 0$ the
central charge of the supersymmetric algebra. It is taken as a
fixed geometrical object, hence the transformation law of
$\mathcal{A}_{s}$ under the gauge transformations generated by the
first class constraint becomes
\begin{equation}\label{eq5}
\begin{aligned}
\delta\mathcal{A}_{r}=D_{r}\epsilon +\{\mathcal{A}_{r},\epsilon\}
\end{aligned}
\end{equation}
It is assumed that it has no transition over $\Sigma$. That is, all
the complicated objects are contained in
$\widehat{\Pi}_{r}$ while $\mathcal{A}_{r}$ is univalued
over $\Sigma$. This is always valid provided we are in the same
principle bundle which is characterized by the winding number $n$.
Under this assumption, which defines a geometrical sector of (\ref{eq1}), the
last term of (\ref{eq1}) becomes zero.

We may now decompose $\mathcal{A}_{r}$ and its canonical conjugate
momenta under the same basis as before:
\begin{equation}\label{eq6}
\begin{aligned}
\mathcal{A}_{r}(\sigma^{1},\sigma^{2},\tau)= & \sum
\mathcal{A}_{r}^{A}(\tau)Y_{A}(\sigma^{1},\sigma^{2})\\
\Pi^{r}(\sigma^{1},\sigma^{2},\tau)= & \sum \sqrt{W}
\Pi^{r,A}(\tau)Y_{A}(\sigma^{1},\sigma^{2})
\end{aligned}
\end{equation}

There is however a main difference between $X^{A}$ and
$\mathcal{A}^{A}_{r}$. It is their transformation law under the
symmetry generated by the first class constraint. To analyze this
point we introduce as in \cite{Marquard} the
structure constants $g^{C}_{AB}$
\begin{equation}\label{eq7}
\begin{aligned}
\{Y_{A},Y_{B}\}=
\frac{2\epsilon^{sr}}{n} D_{r}Y_{A} D_{s}Y_{B}
= g^C_{AB}Y_{C}
\end{aligned}
\end{equation}
Where $Y_{A}$ is a complete orthonormal basis of functions over the spatial part of the
world volume,and $g^C_{AB}$ are the structure constants associated to the group of
area preserving diffeomorphisms in this basis.
That is

\begin{equation}\label{eq8}
\begin{aligned}
 g^C_{AB}=\int d^2\sigma \sqrt{W}\{Y_{A},
 Y_{B}\}Y_{-C}
\end{aligned}
\end{equation}

Where we use the normalization condition

\begin{equation}\label{eq9}
\begin{aligned}
\int d^2\sigma \sqrt{W}{Y_{C}}Y_{B}= \delta_{B+C}
\end{aligned}
\end{equation}

We then have the infinitesimal gauge transformations

\begin{equation}\label{eq10}
\begin{aligned}
\delta X^{C}= & \sum_{A,B}g^C_{AB} \epsilon^{A} X^{B}\\ \delta
\mathcal{A}^{C}_{r}= & -\lambda^{C}_{rA}\epsilon^{A}-
\sum_{A,B}g^C_{AB} \mathcal{A}^{A}_{r} \epsilon^{B}
\end{aligned}
\end{equation}

Where $\lambda^{C}_{rA}$ is defined by
\begin{equation}\label{eq11}
\begin{aligned}
D_{r}Y_{A}=\lambda^{C}_{rA}Y_{C}
\end{aligned}
\end{equation}

We consider $Y_{A}$ to be a complete basis of eigenfunctions of the
operator $D_{r}D_{r}$ . Then we have

\begin{equation}\label{eq12}
\begin{aligned}
D_{r}D_{r}Y_{A}=\omega_{A}Y_{A}
\end{aligned}
\end{equation}

Where no summation in the index $A$ is performed. We assume without loosing
generality that
\begin{equation}\label{eq13}
\begin{aligned}
\overline{Y_{A}}=Y_{-A}
\end{aligned}
\end{equation}
Where $\overline{Y}_{A}$ denotes the complex conjugate to $Y_{A}$.
We notice the following property of the derivatives $D_{r}$
\begin{equation}\label{eq14'}
\begin{aligned}
D_{s}\widehat{\Pi}_{r}=\frac{\widehat{\Pi}^{a}_{s}}
{\sqrt{W}}\partial_{a}\widehat{\Pi}_{r}=
\frac{\epsilon^{ab}}{\sqrt{W}}\partial_{b}
\widehat{\Pi}_{s}\partial_{a}\widehat{\Pi}_{r}
=\{\widehat{\Pi}_{s},\widehat{\Pi}_{r}\}=\frac{n\epsilon_{rs}}{2}
\end{aligned}
\end{equation}
$\widehat{\Pi}_{r}$ may be identified with
 the angles of the compactified directions of the target space.

We may introduce $\widehat{\Pi}_{r}$ as local coordinates over $\Sigma$.We will assume
from now on, $\Sigma$ to be of genus 1, although everything can be extended to arbitrary
genus.
  We then have
\begin{equation}\label{eq13'}
\begin{aligned}
Y_{A}= & \e^{iA_{r}\widehat{\Pi}_{r}}\\ \omega_{A}= &
-\frac{n^{2}}{4}A^{2}_{r}=-\frac{n^{2}}{4}(A^{2}_{1}+ A^2_{2})\\
\lambda^{C}_{rA}= &iA_{s}\frac{n}{2}\epsilon_{sr}
\delta_{A}^{C}\equiv \lambda_{rA}\delta_{A}^{C}
\end{aligned}
\end{equation}
Where $\mathcal{A}_{r},r=1,2$ is a pair of integral numbers associated to
$Y_{A}$.
 The structure functions may then be expressed as

\begin{equation}\label{eq14}
\begin{aligned}
 g^C_{AB}= & \frac{n}{2}(A \times B)\delta^C_{A+B}
\end{aligned}
\end{equation}

and

\begin{equation}\label{eq15}
\begin{aligned}
\lambda^{C}_{rA}= &- i\frac{n}{2}(V_{r}\times A) \delta_{A}^{C}
\end{aligned}
\end{equation}

\begin{equation}\label{eq16}
V_r = \left\{ \begin{array}{cc}
(1,0) & {r=1} \\
(0,1) & {r=2}
\end{array} \right.
\end{equation}

 then

\begin{equation}\label{eq17}
\begin{aligned}
\lambda^{C}_{rA}= & -ig^{C}_{V_r,(A-V_{r})}
\end{aligned}
\end{equation}

and also satisfies

\begin{equation}\label{eq17'}
\begin{aligned}
\frac{2}{n}\epsilon^{sr}\lambda_{rA}\lambda_{sB}d^{C}_{AB}= & \quad g^{C}_{AB}\qquad \textrm{with}\\
 d^{C}_{AB}= & \int d^2\sigma \sqrt{W}Y_{A}Y_{B}\overline{Y}_{C}
\end{aligned}
\end{equation}
With $d^{C}_{AB}$ related to the invariant symmetric three index tensor of
SU(N).

 The Hamiltonian may then be expressed as in \cite{Mari}:

\begin{equation}\label{18}
\begin{aligned}
H = & H_{bosonic} + H_{Fermionic} \\ H_{Bosonic} = & \frac{1}{2}(
P^{0m}P^{0}_{m}+ P^{Am}P^{-A}_{m})+
\frac{1}{4}(g^{C}_{AB}X^{mA}X^{nB})^2 +
\frac{1}{2}(\lambda_{rA}X^{mA}+g^{A}_{BC}\mathcal{A}^{B}_{r}X^{mC})^2 \\ &
+ \frac{1}{2}(\Pi^{rA}\Pi^{-A}_{r}+\Pi^{r0}\Pi^{-0}_{r})
 + \frac{1}{4}[(\lambda_{r}\mathcal{A}_{s}-\lambda{s}\mathcal{A}_{r})^A +
  (g^{A}_{BC}\mathcal{A}^{B}_{r}\mathcal{A}^{C}_{s})]^2+\frac{1}{8}n^2\\
& +\Lambda^{(-A)}( g^{A}_{BC}(X^{mB}P^{C}_{m} + \mathcal{A}^{B}_{r}\Pi^{rC}) + \lambda_{rA}\Pi^{rA}).\\
 \\ H_{Fermionic}= & -g^{A}_{BC}\overline{\Psi}^{(-A)}\gamma_{-}\gamma_{m}X^{mB}\Psi^{C} +
 g^{C}_{AB}\mathcal{A}^{A}_{r}\overline{\psi}^{(-C)}\gamma_{-}\gamma_{r}\Psi^{B} \\  &
  + \lambda_{rB}\overline{\Psi}^{(-B)}\gamma_{-}\gamma_{r}\Psi^{B} -
 g^{A}_{BC}\Lambda^{(-A)}\overline{\Psi}^{B}\gamma_{-}\Psi^{C}.
\end{aligned}
\end{equation}
 Where $( ,)^2$ is understood as:

\begin{equation}\label{eq177}
\begin{aligned}
(\star,\diamond)^2=(\star,\diamond)\overline{(\star,\diamond)}=
(\star,\diamond)^{A}(\star,\diamond)^{-A}
\end{aligned}
\end{equation}

Using the following definitions, $ H_{bosonic}$ can be directly re-expressed in a
simpler way, which may be useful to compare with (\ref{eq1})

\begin{equation}\label{eq19}
\begin{aligned}
\widetilde{\mathcal{D}_{r}}= & \quad\lambda_{r} + [\mathcal{A}_{r},
]\\ \mathcal{F}_{rs}^{A} = & \quad\lambda
_{Ar}\mathcal{A}_{s}^{A}-\lambda_{sA}\mathcal{A}^{A}_{r} +
[\mathcal{A}_{r},\mathcal{A}_{s}]^{A}\qquad \textrm{with}
\\ \lambda_{r}(Y_{A})\equiv & \quad \lambda_{rA}Y_{A}\qquad \textrm{ no summation over index A.}\\
[\star,\diamond]^{A}\equiv & \quad
g^{A}_{BC}\star^{B}_{r}\diamond^{C}_{s}
\end{aligned}
\end{equation}

 Then the bosonic part of the Hamiltonian appears as,

\begin{equation}\label{eq20}
\begin{aligned}
H_{Bosonic} = & \frac{1}{2}P^{0m}P^{0}_{m}+ \frac{1}{2} (P^{Am}P^{-A}_{m})+
\frac{1}{4}[X^{m},X^{n}]^2 +\frac{1}{2}(\widetilde{\mathcal{D}}_{r}X^{mA})^2 +\\ & \frac{1}{8}n^2
 + \frac{1}{2}\Pi^{r0}\Pi^{-0}_{r}+
  \frac{1}{2}\Pi^{rA}\Pi^{-A}_{r} + \frac{1}{4}(\mathcal{F}^{A}_{rs})^2\\
 & + \Lambda^{A}([ X^{m},P_{m}] + \widetilde{\mathcal{D}}_{r}\Pi_{r})^{-A}
\end{aligned}
\end{equation}

Where summation over A index is performed.

\section{The Heisenberg-Weyl Group and the \\
  $N\to\infty$ limit}
We follow in the first part of this section standard results concerning the
Heisenberg-Weyl group \cite{Floratos}. We do so since in the literature there are some minor
misprints that we would like to avoid.

The relevant Hilbert space $H(\Gamma)$ of functions on a torus $\Gamma =\mathcal{C}/L$
of complex modulus $\tau= \tau_{1} + i \tau_{2}$ with integer lattice
$L=\{m_{1}+ \tau m_{2} \mid (m_{1},m_{2}) \in Z \times Z\}$ is defined as the
space of functions of complex argument $z=\sigma_{1} + i\sigma_{2}$

\begin{equation}\label{eq21}
\begin{aligned}
f(z)=\sum_{n \in Z} C_{n}\e^{i\pi(n^{2})+2\pi(inz)}
\end{aligned}
\end{equation}

With the norm

\begin{equation}\label{eq22}
\begin{aligned}
\parallel f \parallel^2 =\int_{\Sigma}d^2\sigma \quad \e^{-2\pi(y^{2})/
\tau_{2}}\mid
f(z)\mid^2
\end{aligned}
\end{equation}

The subspace $H_{N}(\Gamma)$ of $H (\Gamma)$ is defined by the periodicity
condition
\begin{equation}\label{eq23}
\begin{aligned}
C_{n}= C_{n+N}
\end{aligned}
\end{equation}
 for a fixed natural number $N$. In th subspace $H_{N}(\Gamma)$ the discrete
 Heisenberg group with generators P and Q,
\begin{equation}\label{eq24}
\begin{aligned}
Qf(z)= & \sum_{n \in Z}C_{n}\e^{2\pi in/N}\e^{2\pi inz + \pi(in^2 \tau)}\\
Pf(z)= & \sum_{n \in Z}C_{n-1}\e^{2\pi in + \pi(in^2z)}
\end{aligned}
\end{equation}

They satisfy the Weyl relation, \cite{'thooft}:

\begin{equation}\label{eq25}
\begin{aligned}
QP=\kappa PQ \qquad where\qquad \kappa= \exp(2\pi i/N)
\end{aligned}
\end{equation}

The Heisenberg group elements  are defined by

\begin{equation}\label{eq26}
\begin{aligned}
T_{r,s}=N\kappa^{1/2rs}P^{r}Q^{s}
\end{aligned}
\end{equation}

They are SU(N) matrices which satisfy the following relations:
\begin{equation}\label{eq27}
\begin{aligned}
T^{\dagger}_{r,s}= & T_{-r,-s}\\
(T_{r,s})^N= & N^N\e^{i\pi rs(N-1)!} \rm{I}_{N \times N} \\
tr T_{r,s}= & 0\\
T_{r,s}T_{r',s'}= & N \kappa^{1/2(r's-rs')}T_{r+r',s+s'}\\
T_{r+N,s}= & \e^{i\pi s}T_{r,s}\\
T_{r,s+N}= & \e^{i\pi r}T_{r,s}
\end{aligned}
\end{equation}

The SU(N) algebra may be realized in terms of the base
$T_{A}$ with \linebreak $\quad A=(a_1,a_2)=(r,s)$ with $a_1,a_2=0...N$ with
$(0,0)$ excluded. We include $T_{0}=N I_{N \times N}$ to have a complete set of matrices which close under multiplication, \cite{Marquard}

\begin{equation}\label{eq28}
\begin{aligned}
\left[ T_{A},T_{B} \right] =-2iN \sin\left(\frac{(A \times B) \pi}{N}\right)T_{A+B}
\end{aligned}
\end{equation}
 where $[,]$ is simply the commutator, (do not confuse with $[,]$ symbol used in the
last section) the structure constants are then,

\begin{equation}\label{eq29}
\begin{aligned}
f^{C}_{AB} \equiv \frac{1}{N^3} tr([T_{A},T_{B}],T_{-C})=
 -2iN \sin \left(\frac{(A \times B)\pi}{N}\right)\delta^{C}_{A+B}
\end{aligned}
\end{equation}

When $ N\to\infty $ one obtains the Poisson algebra of area preserving
 diffeomorphisms

\begin{equation}\label{eq30}
\begin{aligned}
f_{ABC} \to g_{ABC} \equiv \frac{n}{2}(A \times B)\delta_{A+B-C}
\end{aligned}
\end{equation}

We introduce $\widetilde{\lambda^{B}_{rA}}$  as a particular choice of the structure constants associated to the finite
 group:
\begin{equation}\label{eq300}
\begin{aligned}
\widetilde{\lambda^{B}_{rA}}=-if^{B}_{V_r(A-V_{r})}
\end{aligned}
\end{equation}
with

\begin{equation}\label{eq300'}
\begin{aligned}
\widetilde{\lambda^{B}_{rA}}=\widetilde{\lambda}_{rA}\delta_{A}^{B}
\end{aligned}
\end{equation}

$\widetilde{\lambda}$ converges to
\begin{equation}\label{eq300''}
\begin{aligned}
\frac{ni}{4\pi }\widetilde{\lambda_{rA}}\to\lambda_{rA}=\{\widehat{\Pi}_{r},Y_{A}\}=iA_{s}\frac{n}{2}\epsilon_{rs}
\end{aligned}
\end{equation}

\section{The SU(N) formulation of the Theory}

We may now introduce a SU(N) canonical lagrangian which in the limit $ N\to\infty $
converges to the formulation of section 2 describing the dual of the double
compactified D=11 supermembrane. The coordinates $X^{m}$ as well as the
connection $\mathcal{A}_{r}$ and their canonical conjugate momenta are valued
over the SU(N) algebra. The lagrangian contains unusual terms which indeed are
necessary if  $\mathcal{A}_{r}$ is going to converge to a connection in the
$ N\to\infty $ limit. In fact, a connection of a principle bundle allows to
translate the geometrical objects in the horizontal direction, however in the
SU(N) model all the dependence on the world volume coordinates has been
removed. We then expect some unusual terms which in the $ N\to\infty $ limit
allow that property of the connection 1-form to be recovered.

We are going to consider the Hamiltonian (\ref{18}) in a particular gauge.
Since the first class constraint has been expressed as a generalized Gauss law,
these are several interesting conditions we may impose. We will consider the
gauge condition:
\begin{equation}\label{eq31}
\begin{aligned}
\mathcal{A}_{1}= & \mathcal{A}^{a_1,0}Y_{a_1,0}\\
\mathcal{A}_{2}= & \mathcal{A}^{a_1,a_2}Y_{a_1,a_2} \quad a_2 \neq 0
\end{aligned}
\end{equation}
The SU(N) model we introduce is the following:

\begin{equation}\label{eq32}
\begin{aligned}
H= & tr(\frac{1}{2N^{3}}(P^{0m}T_{0}P^{0}_{m}T_{0}+
\Pi^{r0}T_{0}\Pi^{-0}_{r}T_{0}+(P^{m})^2+ (\Pi)^{2}_{r})
\\& +\frac{n^2}{16\pi^2N^3}[X^{Bm},X^{Cn}]^2+
\frac{n^2}{8\pi^2N^3}(\frac{i}{N}[T_{V_{r}},X^{m}]T_{-V_{r}}- [\mathcal{A}_r,X^{m}])^2\\&
+\frac{n^2}{16\pi^2N^3}([\mathcal{A}_r,\mathcal{A}_s]+
\frac{i}{N}([T_{V_s},\mathcal{A}_r]T_{-V_s}-[T_{V_r},\mathcal{A}_s]T_{-V_r}))^2 + \frac{1}{8}n^2\\&
+\frac{n}{4\pi N^3}
  \Lambda([ X^{m},P_{m}]- \frac{i}{N}[T_{V_r},\Pi_{r}]T_{-V_r} +[ \mathcal{A}_{r},\Pi_{r}])\\
   + &\frac{in}{4\pi N^3}(\overline{\Psi}\gamma_{-}\gamma_{m}\lbrack{X^{m},\Psi}\rbrack
   -\overline{\Psi}\gamma_{-}\gamma_{r}\lbrack{\mathcal{A}_{r},\Psi}\rbrack +
  \Lambda \lbrack{\overline{\Psi}\gamma_{-},\Psi}\rbrack -
   \frac{i}{N} \overline{\Psi}\gamma_{-}\gamma_{r} [T_{V_{r}},\Psi] T_{-V_r}))
\end{aligned}
\end{equation}
subject to
\begin{equation}\label{eq330}
\begin{aligned}
\mathcal{A}_{1}= &\mathcal{A}^{(a_1,0)}_{1}T_{(a_1,0)}, \\
\mathcal{A}_{2}= &\mathcal{A}^{(a_1,a_{2})}_{2} T_{(a_1,a_2)} \quad \textrm{with}\quad a_2\neq0
\end{aligned}
\end{equation}
Where we used the following definitions:
\begin{equation}
\begin{aligned}
X^{m}= & X^{mA}T_{A}\quad\qquad P^{m}=P^{mA}T_{A}\\
\mathcal{A}_r=& \mathcal{A}^{A_{r}} T_{A} \quad\qquad \Pi^{r}=\Pi^{rA}T_{A}\\
[T_{B},T_{C}]= & f^{A}_{BC}T_{A} \\
\end{aligned}
\end{equation}
It may be expressed in the form

\begin{equation}\label{eq321}
\begin{aligned}
H=&tr( \frac{1}{2N^3}(P^{0m}P^{0}_{m}+ (P^{m})^2)+
\frac{1}{2N^3}(\Pi^{0}_{r}\Pi^{0}_{r}+(\Pi^{r})^2)\\ &
\frac{n^2}{16\pi^2N^3}([X^{m},X^{n}]^2 +2(\widehat{\mathcal{D}_{r}}X^{m})^2+
(\widetilde{\mathcal{F}}^{A}_{rs})^2)\ + \frac{1}{8}n^2\\
 & \frac{n}{4\pi N^3}\Lambda(\lbrack X^{m},P_{m}\rbrack +\widehat{\mathcal{D}}_{r}\Pi_{r})\\
   + & \frac{in}{4\pi N^3}\Lambda( \lbrack\overline{\Psi}\gamma_{-},\Psi\rbrack+
   \overline{\Psi}\gamma_{-}\gamma_{m}\lbrack{X^{m},\Psi}\rbrack +
 \widehat{\mathcal{D}}_{r}\overline{\Psi}\gamma_{-}\gamma_{r}\Psi))
\end{aligned}
\end{equation}

where the identification of the terms in (\ref{eq321}) are obvious from
(\ref{eq32}).

Each term of the above Hamiltonian density converges to the corresponding one
in the formulation of the supermembrane of section 2. The
condition (\ref{eq330}) in the SU(N) model also converges to the  gauge fixing condition (\ref{eq31}) of
the supermembrane.
\section{On the Spectrum of the Hamiltonian}

There are two properties of the Hamiltonian of the D=11 supermembrane on a
Minkowski target space \cite{Luscher & Nicolai} which render the spectrum continuous:
\\

i) the existence of local string-like configurations which may even change the
topology of the membrane without changing its energy. This is a property of
the bosonic sector of the supermembrane.\\

ii) Supersymetry.
The Supersymmetry cancels an effective potential coming from the quantization
of the model. It is related to the zero point energy of the harmonic
oscillators which is different from zero in the bosonic case and zero in the
supersymmetric one.
\\

We will show in this section that there are no local string-like configurations
with zero energy density associated to the Hamiltonian of our SU(N) model (\ref{eq32}). It is
important to come back to the global condition which was imposed in order to
obtain the Hamiltonian (see the comment after (\ref{eq5})) of the model under consideration. It was
\begin{equation}\label{eq35}
\begin{aligned}
\int_{\Sigma}\sqrt{W}n\mathcal{^{*}F}=0
\end{aligned}
\end{equation}
 The annihilation of that term (\ref{eq35}) which is
perfectly valid when we formulate our model over a fixed non trivial line
bundle, has important consequences with respect to the
non-existence of the local string-like configurations with zero
energy density. To
analyze this point let us see first what occurs for the compactified
membrane without that assumption.Without the assumption (\ref{eq35}),
 there are local string-like
configurations arising from  following configurations:
\begin{equation}\label{eq36}
\begin{aligned}
X^{m}= & X^{m}(X(\sigma_{1},\sigma_{2}))\\ \mathcal{A}_{r}= &
-\widehat{\Pi}_{r} + f_{r}(X(\sigma_{1},\sigma_{2}))
\end{aligned}
\end{equation}
These configurations depend on an arbitrary uniform map
$X(\sigma_{1},\sigma_{2})$. After some calculations one can show
that the hamiltonian density of (\ref{eq1}) over those configurations becomes zero.
Hence the compactified supermembrane allow local string like spikes with zero energy.
 Let us now discuss the sector of the theory arising from the imposition of the global
condition (\ref{eq35}). In the SU(N) model of section (4), the singular
configurations (\ref{eq36}) do not arise because
$\mathcal{A}_{r}$ is  single valued in distinction to
$\widehat{\Pi}_{r}$ which is necessarily multivalued over $\Sigma$. More precisely,
in order to have zero local energy density, the conditions

\begin{equation}\label{eq37}
\begin{aligned}
\mathcal{D}_{r}X= & 0\\
\mathcal{F}_{rs}= & 0
\end{aligned}
\end{equation}
must be satisfied. They impose severe
restrictions to $X^{A}$ and $\mathcal{A}^{A}_{r}$ which eliminates the possibility of having
the string-like configurations.
In fact, the condition $\mathcal{F}_{rs}=0$ yields
\begin{equation}\label{eq38}
\begin{aligned}
k^{1/2(V_{r}\times A)}\widetilde{\lambda}_{rA}\mathcal{A}^{A}_{s}-
k^{1/2(V_{s}\times A)}\widetilde{\lambda}_{sA}\mathcal{A}^{A}_{r}+f^{A}_{BC}\mathcal{A}^{B}_{r}\mathcal{A}^{C}_{s}=0\\
\end{aligned}
\end{equation}
and using the gauge fixing condition (\ref{eq330}) we obtain, for any $A$
\begin{equation}\label{eq39}
\begin{aligned}
k^{1/2(V_{1}\times A)}N\sin \left(\frac{(V_{1}\times A)\pi}{N}\right)\mathcal{A}^{A}_{2}-
k^{1/2(V_{2}\times A)}N \sin \left(\frac{(V_{2}\times A)\pi}{N}\right)\mathcal{A}^{A}_{1}+\\
N\sin \left(\frac{(b_1V_{1}\times A)\pi}{N}\right)\mathcal{A}^{b_1,0}_{1}\mathcal{A}^{A-b_1V_{1}}_{2}=0\\
\end{aligned}
\end{equation}
where $b_1$ are integers.

In particular for $A=lV_1$ we get
\begin{equation}\label{eq40}
\begin{aligned}
\mathcal{A}^{lV_1}_{1}\equiv \mathcal{A}^{l,0}_{1}=0
\end{aligned}
\end{equation}
hence

\begin{equation}\label{eq41}
\begin{aligned}
\mathcal{A}^{A}_{1}=0
\end{aligned}
\end{equation}

We then obtain from (\ref{eq39}) and the gauge fixing condition
\begin{equation}\label{eq42}
\begin{aligned}
\mathcal{A}^{A}_{2}=0
\end{aligned}
\end{equation}

The condition $ \mathcal{D}_{r}X^{m}=0$ now reduces to
\begin{equation}\label{eq43}
\begin{aligned}
\widetilde{\lambda}_{rA}X^{mA}=0 \qquad r=1,2
\end{aligned}
\end{equation}
That is
\begin{equation}\label{eq44}
\begin{aligned}
N \sin\left(\frac{V_1\times A}{N}\right)\pi X^{mA}=0\\
N \sin \left(\frac{V_2\times A}{N}\right)\pi X^{mA}=0
\end{aligned}
\end{equation}
which yields

\begin{equation}\label{eq44'}
\begin{aligned}
X^{mA}=0
\end{aligned}
\end{equation}
Consecuently, there are no local string like configurations with
 zero energy density for the SU(N) model of section 4.

\section{Conclusions.}

We proposed a model described by SU(N) algebra valued geometrical
objects which in the limit converges to the dual of the double
compactified D=11 supermembrane with nontrivial winding.
 It describes a supermembrane with fixed winding $n \neq 0$ on a target space
  $M_{9}\times S^1 \times S^1$. We showed explicitly the existence of local
  string-like spikes in the general formulation of compactified supermembranes,
   in agreement with \cite{Peeters & Plefka}. We then proved that in the
    proposed SU(N) model for supermembranes with fixed winding, which is only
     one sector of the full theory, there are no string like spikes and hence
      this sector should have discrete spectrum. We will analyzed in more detail
  the properties of the spectrum elsewhere. It is important to remark that this
  sector is described by a global condition which eliminates completely the
  local string like spikes.
\gap

{\bf Acknowledgments} One of the authors$^a$ would like to thank to
 N. Hatcher and J. Ovalle for helpful conversations. We would also like
  to thank King´s College group for kind hospitality while
 part of this work was done.

\newpage

\ 
\end{document}